\documentclass[aps, prb, amsmath, amssymb, reprint,
               superscriptaddress, showkeys]{revtex4-1}
\pdfoutput=1
\usepackage{color}
\usepackage{xcolor}

\usepackage{amssymb,amsmath}

\usepackage{graphicx}

\usepackage{braket}

\usepackage[pdftex,
            pdfauthor={Vanessa Junk, Phillipp Reck, Cosimo Gorini, Klaus Richter},
            pdftitle={Floquet oscillations}]{hyperref}

\newcommand{\bb}[1]{\mathbf{#1}}     
\newcommand{\bs}[1]{\boldsymbol{#1}} 

\newcommand{\rmi}{\textrm{i}} 
\newcommand{\rmd}{\textrm{d}} 
\newcommand{\rme}{\textrm{e}} 

\newcommand{\ie}{{\it i.e.~}}	
\newcommand{\eg}{{\it e.g.~}} 	

\def\?#1{}

\newcommand{\floqueteigenvalues}{\ref{suppeq:floqueteigenvalues}}

\begin{document}
%
%
\title{Floquet oscillations in periodically driven Dirac systems }
\newcommand{\RegensburgUniversity}{Institut f\"ur Theoretische Physik,
  Universit\"at Regensburg, D-93040 Regensburg, Germany}
\author{Vanessa Junk}
\affiliation{\RegensburgUniversity}
\author{Phillipp Reck}
\affiliation{\RegensburgUniversity}
\author{Cosimo Gorini}
\affiliation{\RegensburgUniversity}
\author{Klaus Richter}
\email{klaus.richter@physik.uni-regensburg.de}
\affiliation{\RegensburgUniversity}



\begin{abstract}
Electrons in a lattice exhibit time-periodic motion, known as Bloch oscillation,
when subject to an additional static electric field.
Here we show that a corresponding dynamics can occur upon replacing the spatially periodic 
potential by a time-periodic driving: 
Floquet oscillations of charge carriers in a spatially homogeneous system.
The time lattice of the driving gives rise to Floquet bands that take on the role of the usual Bloch bands.
For two different drivings (harmonic driving and periodic kicking through pulses) of systems with linear dispersion we demonstrate the existence of such oscillations, both by directly propagating wave packets and based on a complementary Floquet analysis.
The Floquet oscillations feature richer oscillation patterns 
than their Bloch counterpart and enable the imaging of Floquet bands.
Moreover, their period can be directly tuned through the driving frequency.
Such oscillations should be experimentally observable in effective Dirac systems, 
such as graphene, when illuminated with circularly polarized light.    
\end{abstract}

\maketitle

\section{Introduction}

In the early days of quantum mechanics, F.~Bloch and C.~Zener \cite{Bloch1929, Zener1934} predicted 
that electrons in a periodic potential, when  accelerated by a constant external electric field,
perform a time-periodic motion, by now well known as Bloch oscillation \cite{wacker2002}.
It took about 60 years until this phenomenon was observed in biased semiconductor superlattices
 \cite{keldysh1963, feldmann1992, bofirstmeasurement3, ignatov1994}. 
Since then Bloch oscillations or analogs of them have been found in various systems 
ranging from cold atom gases~\cite{bocoldatoms, bocoldatoms2} to classical 
optical \cite{bowaveguide, bowaveguides2} and acoustic waves \cite{boacoustic}, to name a few.
In 2014, Bloch oscillations due to the crystal lattice of a biased bulk semiconductor
were eventually observed~\cite{BlochcrystalHuber}. 

In the meantime, scientific interest in tuning Bloch bands by means of external time-periodic 
driving has rapidly grown, especially since the proposal of so-called Floquet topological 
insulators~\cite{FloquetTIGalitski} demonstrating the powerful influence external driving 
can exert on the properties of a crystal.
Recent experiments also showed that Floquet band engineering allows for
switching Bloch oscillations on and off \cite{FloquetMeasurementFujiwara}. 
Moreover, additional driving can immensely increase the amplitude of conventional 
Bloch oscillations, giving rise to ``super'' Bloch 
oscillations~\cite{SuperBlochNature, SuperBlochTransport, Holthaus_super_Bloch, SuperBlochTheory}. 


Here we propose to consider the opposite limit of a time-periodically driven system without any spatial lattice,
but still subject to a constant external electric field. We demonstrate that, most notably, still 
spatially periodic motion of the charge carriers can appear. We call this type of dynamics
{\em Floquet oscillations} since 
they arise from the periodic repetitions of Floquet quasi-energy bands.
So far very few works have addressed Bloch-type oscillations in the absence of an external lattice.
One interesting prediction refers to Bloch oscillations of light, \ie frequency oscillations of 
photons~\cite{BlochLightYuan}. Further Bloch-type oscillations were predicted theoretically for 
interacting 1d spinor gases~\cite{gangardt2009} and recently observed in an atomic Bose 
liquid~\cite{Meinert945}. 
In these settings interactions lead to the dynamical formation of periodic structures,
which can yield oscillations \`a la Bloch.  They do not, however, involve external drivings, and hence
are of different nature than the Floquet oscillations predicted here.

Specifically, we show that such periodic modes can emerge in spatially uniform systems governed by effective
Dirac Hamiltonians, where the linear dispersion converts the energy periodicity of the Floquet 
spectrum into approximately $k$-periodic bands. 
Most notably, Floquet oscillations are a quantum phenomenon distinctly different from the classical oscillatory motion of a charge in an ac field. Instead of following the driving frequency, they exhibit a frequency inversely proportional to it.
Moreover, Floquet oscillations offer a possibility to directly image 
 the Floquet quasi-band structure.  
Interestingly, they additionally show zitterbewegung features.
We support our predictions by numerical calculations for two experimentally relevant prototypes of
external driving, a periodic pulse sequence and circularly polarized radiation.


 \section{General concept of Floquet oscillations}


\begin{figure}
\centering
\includegraphics[width=0.9\columnwidth]{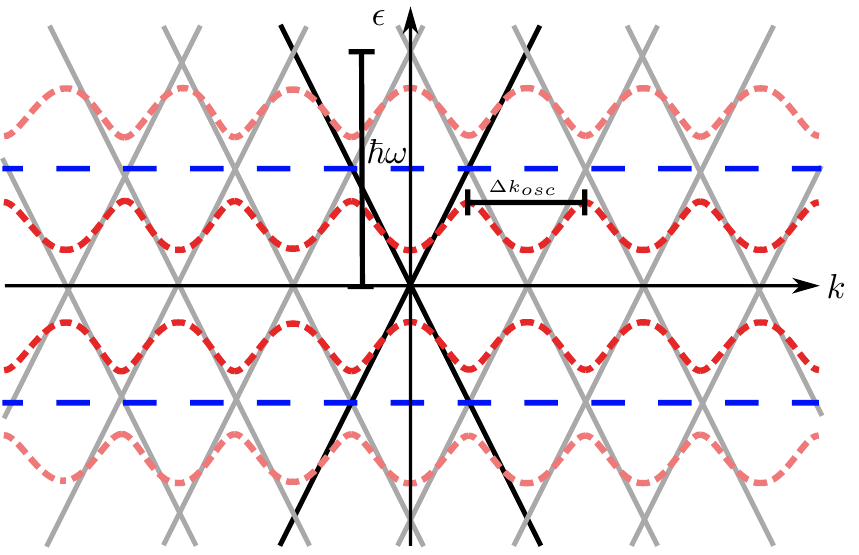}
\caption{Simplified scheme of the Floquet band structure $\epsilon_\pm(k)$ of a Dirac system subject to a 
time-periodic potential.
In Floquet theory the driving (at frequency $\omega$) introduces replicas of the original Dirac cone 
$E_\pm=\pm \hbar v_F k$ (black), vertically shifted in energy by multiples of $\hbar \omega$ (thin gray).
The blue dashed lines mark the first energy unit cell. 
For finite driving strength the bands hybridize and gaps open at the crossings, leading 
to distinct (approximately) $k$-periodic Floquet bands (red dashed) 
whose local maxima are $\Delta k_{osc}$ apart.
}
\label{floquetbandslinear}
\end{figure}


Consider a system $H_0(\bb k)$ (with momentum operator $\bb k$) subject to a
time-periodic driving $V_T(t)$ with period $T$ and frequency $\omega = 2\pi/T$ described by the Hamiltonian
\begin{equation}
H(\bb k,t)=H_0(\bb k) + V_T(t) = H(\bb k,t+T) \, .
\label{eq:floquet-general}
\end{equation}
Via Floquet theory \cite{Floquet1883, floquetince, floquethill} the problem is transformed 
to finding the quasi-energy eigenvalues $\epsilon(\bb k)$ of the Floquet Hamiltonian
$\mathcal{H}_F(\bb k)=H(\bb k,t)-\rmi \hbar \frac{\partial}{\partial t}$. 
The quasi-energies $\epsilon(\bb k)$ extend to infinity in $\bb k$-space in absence of a spatial lattice.
However, they are periodic in quasi-energy $\epsilon\in [-\hbar\omega/2,\hbar\omega/2]$ forming a sequence of Floquet replica
which are the analogue of usual Bloch bands -- the latter being periodic in quasi-momentum $k\in [-\pi/a,\pi/a]$ due to spatial periodicity with lattice constant $a$.

For the undriven system, we take an (effective) two-dimensional Dirac Hamiltonian
\begin{equation}
H_0(\bb k)=\hbar v_F \bb k \cdot \boldsymbol{\sigma}, 
\label{DiracHamiltonian}
\end{equation}
with constant Fermi velocity  $v_F$ and $\boldsymbol{\sigma}=(\sigma_x, \sigma_y)$ the vector of Pauli matrices.
The spectrum of $H_0(\bb k)$ is composed of two energy branches 
$E_\pm = \pm \hbar v_F k$ with $k = |\bb k|$. 
The Floquet bands $\epsilon_\pm$ that emerge when adding the time-periodic driving $V_T(t)$ to the system are sketched in Fig.~\ref{floquetbandslinear}.
Due to the radial symmetry of the bands, we only show a cut along $k_y = 0$. In the limit of infinitely small driving $V_T(t)$, the bare Dirac cone $E_\pm$ (black) is accompanied by a mesh of intersecting replica (gray) that are shifted in energy by multiples of $\hbar \omega$.
The blue dashed lines mark the corresponding first Brillouin zone (BZ) in energy.
For finite $V_T(t)$ band gaps open at replica crossings and separated Floquet bands emerge
(red dashed curves, see Appendix \ref{suppFloquet} for an introduction to the relevant aspects of Floquet theory). 
The band gaps sketched in Fig.~\ref{floquetbandslinear} emerge since different replica are coupled. Under the influence of the drift field $E$, a particle follows the band adiabatically along $k$ and successively absorbs/emits an increasing number $n$ of photons $\hbar \omega$ at each avoided crossing between two replica of the (undriven) Dirac cone with relative energy shift $n\hbar\omega$.
Hence, the farther away from $k=0$, the more photons are needed to bridge the energy difference required for adiabatic dynamics associated with Floquet oscillations. 
As we will show below, time-periodic potentials $V_T$ with pronounced higher-order (in $\omega$) Fourier components will correspondingly open band gaps farther away from $k=0$ compared to, {\it e.g.}, single frequency harmonic driving. Due to the underlying linear (Dirac) dispersion, the Floquet bands are approximately $k$-periodic, implying particularly pronounced Floquet oscillations with a well-defined frequency, in
close analogy to Bloch oscillations
\footnote{General nonlinear dispersions $\epsilon_\pm(k)$, \eg two hyperbolic branches, give rise to sequences of intersections that are not equidistant in $k$ and thereby deny a constant Floquet oscillation period.}.

To gain insight into the Floquet oscillation dynamics let us 
be definite and consider the time evolution of a wave packet (WP) of 
Dirac electrons under the influence of an additional, constant electric 
field
$\bb E = E \bb e_x$.
Here we consider the 1d motion along the field direction, generalizations to higher
dimensions are straight forward.  Due to the drift potential 
\begin{equation}
V (x) = - eE x,\ e>0,
\label{eq:Efield}
\end{equation}
the WP is accelerated and its initial wave vector $k_i$ changes linearly in time
\cite{Holthaus_super_Bloch}: 
 \begin{equation}
k(t)= k_i - (e E/\hbar) t \, .
\label{relation_time_k}
 \end{equation}
For ordinary Bloch bands, the BZ with period $\Delta k_{osc}\!=\! 2\pi/a$ is traversed in the time 
\begin{equation}
T_B = 2\pi/(a|eE|\hbar).
\label{eq:T-Bloch}
\end{equation}   
While crossing the BZ, a charge carrier changes its velocity according to the change in slope
of the $k$-periodic band structure, leading to a Bloch oscillation in $k(t)$ with  
frequency $\omega_B=2\pi/T_B$. 

For Floquet systems the velocity operator is given by $\hat{v} = \rmd \mathcal{H}_F/ \rmd k$~\cite{Holthaus_super_Bloch}.
The diagonal terms of a WP's velocity read 
$\langle \hat{v}_{\alpha \alpha} \rangle \!=\! v_{\alpha \alpha}\!=\! \rmd \epsilon_\alpha/ \rmd k$, with $\alpha\!=\!\pm$,
and the corresponding position expectation value $\langle \hat{x}_{\alpha \alpha} \rangle = x_{\alpha \alpha}$ 
is obtained by time integration of $v_{\alpha \alpha}$.
Using Eq.~\eqref{relation_time_k} this can be substituted by an integration over $k$ leading to
\begin{equation}
x_{\alpha \alpha}(k(t)) 
= \frac{\hbar}{e E}\left[\epsilon_\alpha(k(t)) - \epsilon_\alpha(k_i)\right] \, .
\label{relation_x_v}
\end{equation}
These diagonal contributions to $\langle \hat{x} \rangle (t)$ encode features of the
Floquet band structure into the WP position.  In particular, analogously to conventional 
Bloch oscillations, the WP is expected to perform oscillations for Floquet bands similar to the ones sketched in Fig.~\ref{floquetbandslinear} (red dashed lines). 

During one (Floquet) oscillation $k$ changes by the period $\Delta k_{osc}$ of the Floquet bands
(Fig.~\ref{floquetbandslinear}). 
Hence, Eq.~(\ref{relation_time_k}) implies the corresponding period
$T_F =  (\hbar/|eE|) \Delta k_{osc}$.
Due to the linear dispersion $E_\pm(k)=\pm \hbar v_F k$, we have
$\hbar v_F\Delta k_{osc} = \hbar \omega$ 
so that the Floquet oscillation period reads
\begin{equation}
T_F =  \frac{\hbar \omega}{v_F |e E|} \, .
\label{eq:T-Floquet}
\end{equation}  
$T_F$ is proportional to the {\em inverse} period $1/T=\omega/(2\pi)$ 
of the driving in Eq.~(\ref{eq:floquet-general}).
While its analogue, the Bloch period $T_B$, Eq.~(\ref{eq:T-Bloch}), 
is determined by the inverse \mbox{(super-)lattice} constant $1/a$, usually fixed in experiment, 
$T_F$ can be tuned through $\omega$ in a range such that $T_F > T$.
 

\section{Results for representative driving protocols}

\begin{figure}
\centering
\includegraphics[width=0.85\columnwidth]{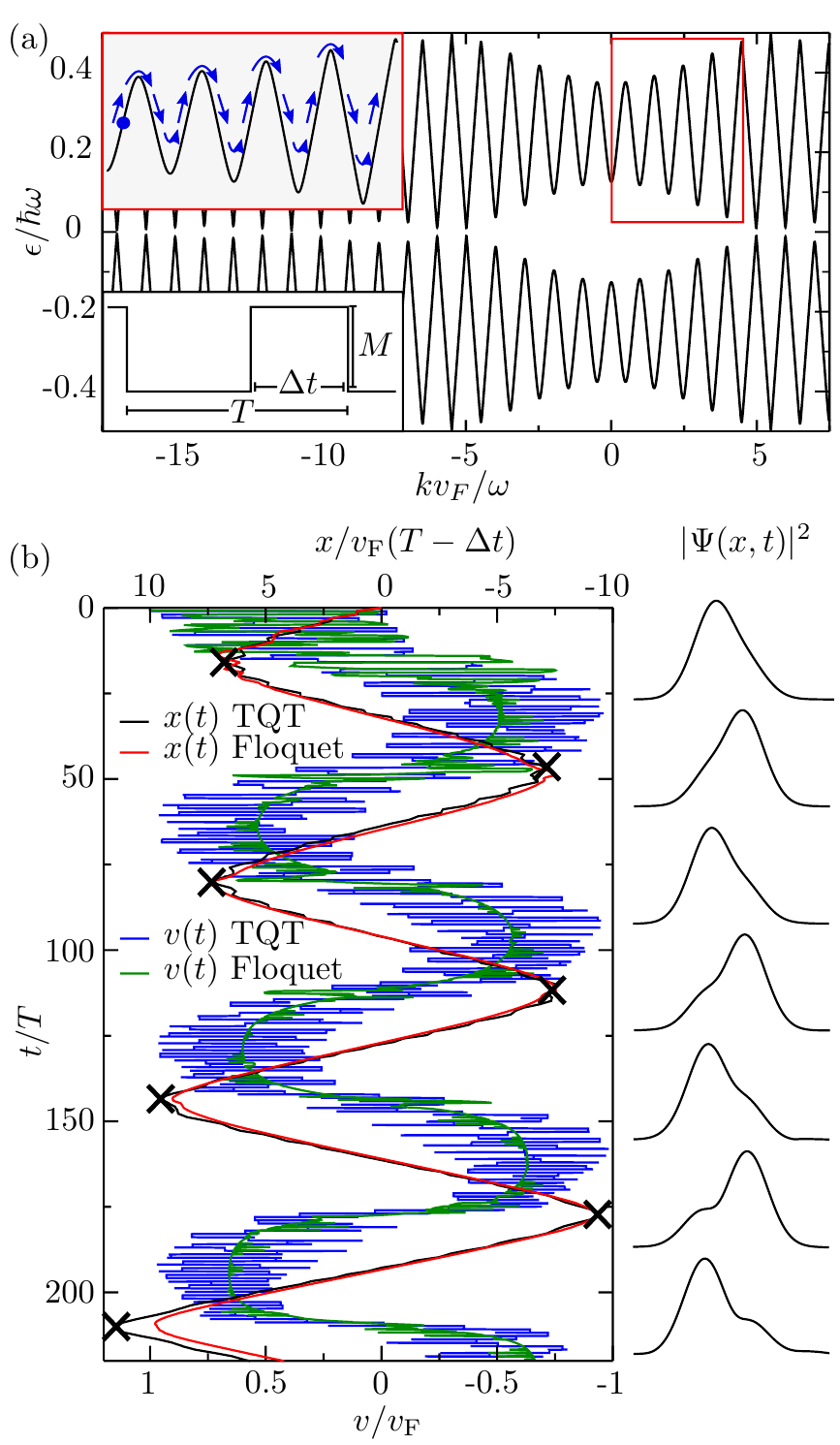}
 \caption{
Floquet oscillations in a driven, spatially homogeneous Dirac system.
(a) Floquet bands from a periodically opened mass gap (see Eq.~(\ref{potentialmass})
and lower left inset for the pulse shape) with band gaps at the intersections
of the unperturbed Dirac cones and their replicas (as sketched in Fig.~\ref{floquetbandslinear}).
Two separated bands emerge in the energy window $\hbar \omega$. 
The motion of a WP (with initial position indicated by the blue dot) due to a static electric field, Eq.~\eqref{eq:Efield},  
is sketched by the blue arrows in the zoomed-in region framed in red.
(b) Left panel: Simulations of the WP position and velocity expectation values
based on direct time evolution (TQT) and Floquet theory (see text for details).
The numerical results confirm the Floquet oscillations mirroring the characteristic shape of the Floquet band structure.
They also show zitterbewegung signatures.
Right panel: Snapshots of the WP in position space taken at times marked as black crosses in the left panel. 
}
\label{massgap}
\end{figure}

In the following we consider two representative types of driving well suited to generate Floquet 
oscillations: a periodic sequence of short pulses and a circularly polarized light field.

\subsection{Short pulse sequence}

For the first driving protocol we use 
\begin{equation}
V_T(t) = \sum_{l=1}^{\infty} \Theta (t-(lT-\Delta t)) \Theta (lT-t) \  M\sigma_z \, 
\label{potentialmass} 
\end{equation} 
in Eq.~(\ref{eq:floquet-general})
(with  Heaviside function $\Theta$ and Pauli matrix $\sigma_z$).
This pulse train periodically couples the two branches of the Dirac Hamiltonian $H_0({\bb k})$, Eq.~\eqref{DiracHamiltonian}, 
by opening a mass gap of strength $M$ and duration $\Delta t$, see lower inset in Fig.~\ref{massgap}(a). 
The resulting Floquet spectrum can be tuned either by $M$ or by the ratio $\Delta t/T$.
To be definite we choose the normalized pulse strength $M/\hbar \omega = 4.4/\pi$ and $\Delta t/T = 0.09$.
The resulting Floquet band structure for this representative set of parameters is shown in 
Fig.~\ref{massgap}(a). 
The driving opens a sequence of gaps around $\epsilon\!=\!0$ 
at the intersections of the original Dirac spectrum and its replicas.
A detailed analysis of these Floquet bands and the $k$-dependence of 
the gaps is given in Appendix \ref{suppBands}.

A WP with initial quasi-energy and wave number as marked by the blue dot in the zoomed in 
area in Fig.~\ref{massgap}(a) (red box, $\bar{k}_i v_F/\omega = 0.25$) will undergo Floquet oscillations (blue arrows) when driven 
through the bands by a static electric field $E<0$, Eq.~\eqref{eq:Efield}, such that $k$ increases in time, Eq.~\eqref{relation_time_k}. 
Notably, since the Floquet band maxima and hence the band width vary over $k$-space, 
the resulting Floquet oscillation is expected to change its amplitude but not its frequency. 
In our simulations we choose the field strength $E$ such that, according to Eq.~(\ref{eq:T-Floquet}), 
$T_F/T = 2 \pi \hbar / T^2 v_F |eE| \simeq 20.8 \pi$.
We took initially Gaussian WPs of the form
\begin{equation}
\tilde\Psi(k,0) = \sqrt{\frac{1}{\sqrt{\pi}\Delta k}} \exp 
\left( -\frac{1}{2\Delta k^2} (k\!-\!\bar{k}_i)^2 \right) \cdot \begin{pmatrix}
1 \\ 1
\end{pmatrix},  
\label{initialGaussian}
\end{equation}
with width $\Delta k$ and initial central mode $\bar{k}_i$.
We employ two complementary approaches to compute and analyze Floquet oscillations:
Floquet theory and direct time-integration of the full time-dependent effective Dirac equation
including the $\bf E$-field.

To compute the WP velocity within Floquet theory we start from Ehrenfest's theorem 
\begin{equation}
v(t) =\langle \hat{v}(t) \rangle = \frac{i}{\hbar} \langle \Psi(t) | \left[ H(t), \hat{x} \right] | \Psi (t) \rangle,
\end{equation}
where $\left[ H(t), \hat{x} \right] = -i\hbar v_F \sigma_x$ for the Dirac case.
$| \Psi(t)\rangle$ 
is obtained via the time-evolution operator of a Floquet system \cite{floquethaenggi}
that for a single $k$-mode reads
\begin{equation}
U_k(t, 0) = \sum_{\alpha=\pm} \exp \left(-\frac{i}{\hbar} \epsilon_\alpha (k) 
t \right) | \phi_{\alpha,k} (t) \rangle \langle \phi_{\alpha,k} (0) | \, .
\label{timeevolutionfloquet}
\end{equation}
Here $\epsilon_\alpha $ are the Floquet quasi-energies and 
$| \phi_{\alpha,k} (t) \rangle$ the corresponding eigenstates of ${\cal H}_F$, 
including replicas $n\hbar \omega$ (see Eq.~(\floqueteigenvalues)),
each consisting of two branches $\alpha\!=\!\pm$ from the linear dispersion.
The additional electric field induces a linear change of $k$, which we account for 
by adjusting $k(t)$  according to Eq.~\eqref{relation_time_k}. 
Applying $U_k(t, 0)$ to an initial (WP) state
\begin{equation}
| \Psi(0)\rangle = \sum_{k_i} c_{\alpha, k_i}|\phi_{\alpha, k_i}(0)\rangle,
\end{equation} 
where
$|c_{\alpha, k_i}|^2 = |\langle \phi_{\alpha, k_i}(0) | \Psi(0)\rangle |^2$ 
describes the initial occupation of branch $\alpha$, and plugging 
Eq.~\eqref{timeevolutionfloquet} into Ehrenfest's theorem gives
\begin{align}
v(t) =& v_F \sum_{k_i}
\sum_{\alpha, \beta=\pm} c_{\alpha, k_i}^\ast c_{\beta, k_i}
\langle \phi_{\alpha, k(t)}(t)| \sigma_x |\phi_{\beta, k(t)} (t) \rangle \nonumber \\
& \qquad \quad \times \exp 
\left(- \frac{i}{\hbar} \left[ \epsilon_\beta \left(k(t)\right)- 
\epsilon_\alpha \left(k(t)\right] \right)t \right) \, .
\label{floquetvelocity}
\end{align}
Here $k(t)$ is given by
Eq.~\eqref{relation_time_k}.
The occupation $|c_{\alpha, k_i}|^2$ is time-independent as long as different Floquet bands
are far enough apart for Landau-Zener interband transitions 
\cite{landau1932theory, Zener696, landau1932phase, stueckelberg1933, BlochZenerBreid, krueckl2012} 
to be neglected. This is the case for the time scales $t\le 200T$ considered below.

For the periodic pulse sequence, Eq.~({\ref{potentialmass}),
we numerically compute $v(t)$ by means of Eq.~\eqref{floquetvelocity} and  
$x(t)=\int_0^t \rmd t^\prime v(t^\prime)$.  Due to the rectangular pulse shape 
we must include up to $n=500$ Floquet modes to achieve sufficient convergence.
The results are shown in the 
left panel of Fig.~\ref{massgap}(b) as red and green lines for $x(t)$ and  $v(t)$, respectively. 
They indeed show distinct Floquet oscillations, as predicted, with period $T_F \simeq 20.5\pi T$, 
matching the expected value $T_F \simeq 20.8\pi T$ from Eq.~(\ref{eq:T-Floquet}). The off-diagonal velocity term ($\alpha \neq \beta$) in Eq.~(\ref{floquetvelocity}) encodes the interference 
of states living in different Floquet bands, giving rise to an additional feature,
zitterbewegung \cite{schroedinger1930zitterbewegung, zitterbewegungrusinreview} 
caused by the Floquet Dirac band structure (see Appendix \ref{suppZB}).

Our second, complementary method to compute Floquet oscillations is based on the WP
propagation algorithm ``Time-dependent Quantum Transport'' (TQT) \cite{phdkrueckl}, see also Appendix \ref{suppTQT}. 
The WP is discretized on a rectangular grid and the time-evolution operator for the
full Hamiltonian including the drift field, 
$ H(t)\!=\! H_0(\bb k) \!+\! V_T(t) \!-\! E x $,
is computed. Then the Lanczos method \cite{lanczos} is used to evaluate the action of the
time-evolution operator on the WP and to compute $\Psi(x,t)$. The time-dependent position and velocity expectation values are then calculated through
\begin{equation}
x(t) = \langle \hat{x}(t) \rangle = \int  |\Psi(x, t)|^2 x \ \rmd x
\label{eq:x_TQT} 
\end{equation}
and
\begin{equation}
v(t) =  \langle \hat{v}(t) \rangle = \frac{\rmd}{\rmd t} \langle \hat{x}(t) \rangle.
\label{eq:v_TQT}
\end{equation}
The resulting TQT data is shown in Fig.~\ref{massgap}(b), left panel, 
as black and blue curves for $x(t)$ and  $v(t)$, respectively. 
They approximately coincide with those computed within Floquet theory, also showing
zitterbewegung on top of the Floquet oscillations.
Moreover, the WP position $x(t)$ computed with TQT precisely reflects characteristic 
features of the Floquet quasi bands, shown in the red box in Fig.~\ref{massgap}(a), namely the 
increasing amplitude and the sharpening of the maxima and minima although TQT directly integrates the time-dependent Schr\"odinger equation without using Floquet formalism.

While the static electric drift field enters into the full Hamiltonian governing the numerically
exact TQT time evolution, within the Floquet approach its effect is included via the acceleration 
theorem \eqref{relation_time_k} into the time evolution, 
Eqs.~(\ref{timeevolutionfloquet}, \ref{floquetvelocity}). The latter approximation, together
with residual numerical errors from the cutoff in the Floquet quantum number,
could explain the slight deviations between Floquet and TQT data in the left panel
of Fig.~\ref{massgap}(b).

Finally, in the right panel of Fig.~\ref{massgap}(b) we present snapshots of the absolute square $|\Psi(x,t)|^2$ 
taken at the turning points (marked as black crosses) of the red curve in the left panel. They
show clear-cut Floquet oscillations of the full WP in configuration space generated
for the setting of a periodic pulse sequence.

\subsection{Circularly polarized light}

The experimental realization of Floquet oscillations could be easier to achieve in an alternative setup,
employing circularly polarized light as periodic driving.
The associated vector potential $\bb A$ enters (linearly) the Dirac Hamiltonian (\ref{eq:floquet-general}) 
via the minimal coupling
\begin{equation}
V_T(t) = \bb A(t) \cdot \bs\sigma = A \begin{pmatrix}
\cos (\omega t) \\ \sin (\omega t) 
\end{pmatrix} \cdot \bs\sigma.
\label{eq:lightfield}
\end{equation}
The Floquet quasi-bands of graphene illuminated by circularly polarized light have already been studied 
extensively \cite{HallgrapheneOka, GrapheneInsulatorKibis, GrapheneTHzWu, GrapheneLaserSergey}. 
Recently, also transport \cite{IrradiatedGrapheneballistic} and topological 
\cite{GrapheneLightBerry, TopTransitionGrapheneGloria, GrapheneFTIBalseiro, HierachyGapsHoneycombUsaj} 
properties were investigated.  Instead, here we focus on generating Floquet oscillations for realistic
experimental parameters.  To be closer to measurements, we explicitly treat the 2d case with an initial, 
radially symmetric Gaussian WP analogous to Eq.~\eqref{initialGaussian}, with $k$ and $\bar{k}_i$ replaced by $\bb k$ and $\bar{\bb k}_i = (\bar{k}_i , 0)$.
Using again TQT we simulate the WP dynamics (see Fig.~\ref{x-y_WP_circ})
in presence of the electric field $\bb E = E \bb e_x$, where $T_F/T = 2\pi \hbar / T^2 v_F |eE| \simeq 140$.

\begin{figure}
\centering
\includegraphics[width=\columnwidth]{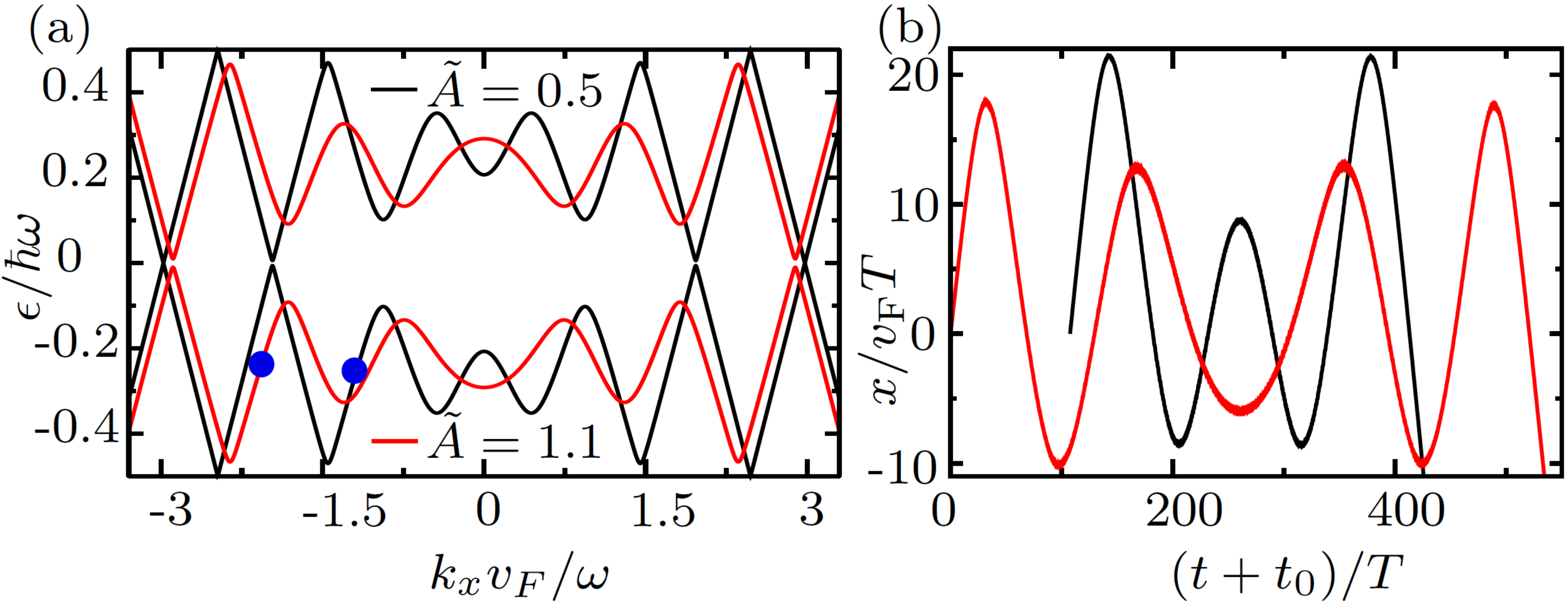}
\caption{(a) Floquet bands of a Dirac system illuminated with circularly 
polarized light with scaled amplitude $\tilde{A} =$ 0.5 (black curves) and 1.1 (red curves).
(b) Floquet oscillations: Mean position of a WP driven through the bands shown 
in (a) starting with momenta $\bar{k}_i$ marked as blue dots in (a). The initial time $t_0$ for the black curve is shifted to $t_0/T=107.5$ to highlight the close connection
to the Floquet band structure of panel (a). Additional oscillations due to the momentum change caused by the circulating electric field of the pulse are not resovled on this timescale.}
\label{circpollight}
\end{figure}

We calculate the Floquet band structure and the WP dynamics for two different dimensionless 
driving strengths $\tilde{A}=v_F e A/(\hbar \omega)$
to demonstrate the field amplitude influence on the Floquet oscillation frequency. 
Figure \ref{circpollight}(a) shows the radially symmetric Floquet band structure along the direction of the electric field. 
For small enough driving strengths, the distance of local band minima is independent of $k$ (black). For larger driving amplitudes, multiple-photon processes significantly alter the formation of band gaps. 





As displayed in Fig.~\ref{circpollight}(b), our TQT simulations of the position expectation values, Eq.~\eqref{eq:x_TQT}, of 
two WPs with scaled initial momenta $\bar{k}_i v_F/\omega =-2.07$ and $-1.22$
for $\tilde{A} =$ 1.1 and 0.5, respectively (marked by blue dots in panel (a)) clearly show Floquet oscillations, nicely reflecting the shape of the underlying
Floquet band structure as expected. Since the gaps between unperturbed Dirac cones open in a smaller $k$-range than for the periodic
pulse train \eqref{potentialmass},
there are less cycles of regular Floquet oscillations. At longer times
Landau-Zener transitions to neighboring Floquet bands substantially alter the WP motion.
Nevertheless, Fig.~\ref{circpollight} shows Floquet oscillations with 4 full periods. In the famous experiments of Bloch oscillations in superlattices \cite{feldmann1992, bofirstmeasurement3, ignatov1994} and bulk semi-conductors \cite{BlochcrystalHuber}, their detection was possible even though only $1-3$ periods could be achieved.
Moreover, oscillations involving many periods exist for the case of periodic kicking which can be experimentally realized through laser pulse trains.

In analogy to Fig.~\ref{massgap}(b), Fig.~\ref{x-y_WP_circ}(a) shows the position expectation value, Eq.~\eqref{eq:x_TQT}, along the electric field, Eq.~\eqref{eq:Efield}, and the 2d shape of the WP. Here we choose $\bar{k}_i v_F/\omega = -0.64$ and $T_F/T = 2\pi\hbar / T^2 v_F |eE| \simeq 14$ without shifting the initial time $t_0$ and including orange crosses to mark when the snapshots of the WP are taken. The snapshots displayed on the right show the absolute square of the WP in 2d space and the oscillation of its center of mass around $x=0$. Note that according to Eq.~\eqref{eq:T-Floquet} $T_F$ is inversely proportional to $E$, therefore in Fig.~\ref{x-y_WP_circ} the timescale of the Floquet oscillations is a factor of 10 smaller than in Fig.~\ref{circpollight}. Here, the oscillations induced by the circulating electric field, Eq.~\eqref{eq:lightfield}, are resolved on top of the slower Floquet oscillations. In Fig.~\ref{x-y_WP_circ}(b), we plot the trajectory of the WP's center of mass in the $x$-$y$-plane. The red curve is meant to serve as a guide for the eye. In the trajectory, one can recognize the Floquet oscillations but also some additional features. The smaller oscillations have the same origin as the tiny oscillations on top of the comparably slow Floquet oscillations in Fig.~\ref{x-y_WP_circ}(a): They arise from the momentum change due to the circulating electric field of the pulse, Eq.~\eqref{eq:lightfield}, and are modulated by zitterbewegung. The drift of the WP in $y$-direction however has a less intuitive explanation. We assume that it is caused by the anomalous velocity, \ie the Berry curvature of the Floquet bands, since it is orthogonal to the electric driving field. This assumption will be tested in future studies.

\begin{figure}
\vspace*{0.4cm}
\centering
\fontsize{10pt}{0pt}\selectfont
\def\svgwidth{1\columnwidth}
\includegraphics[width=\columnwidth]{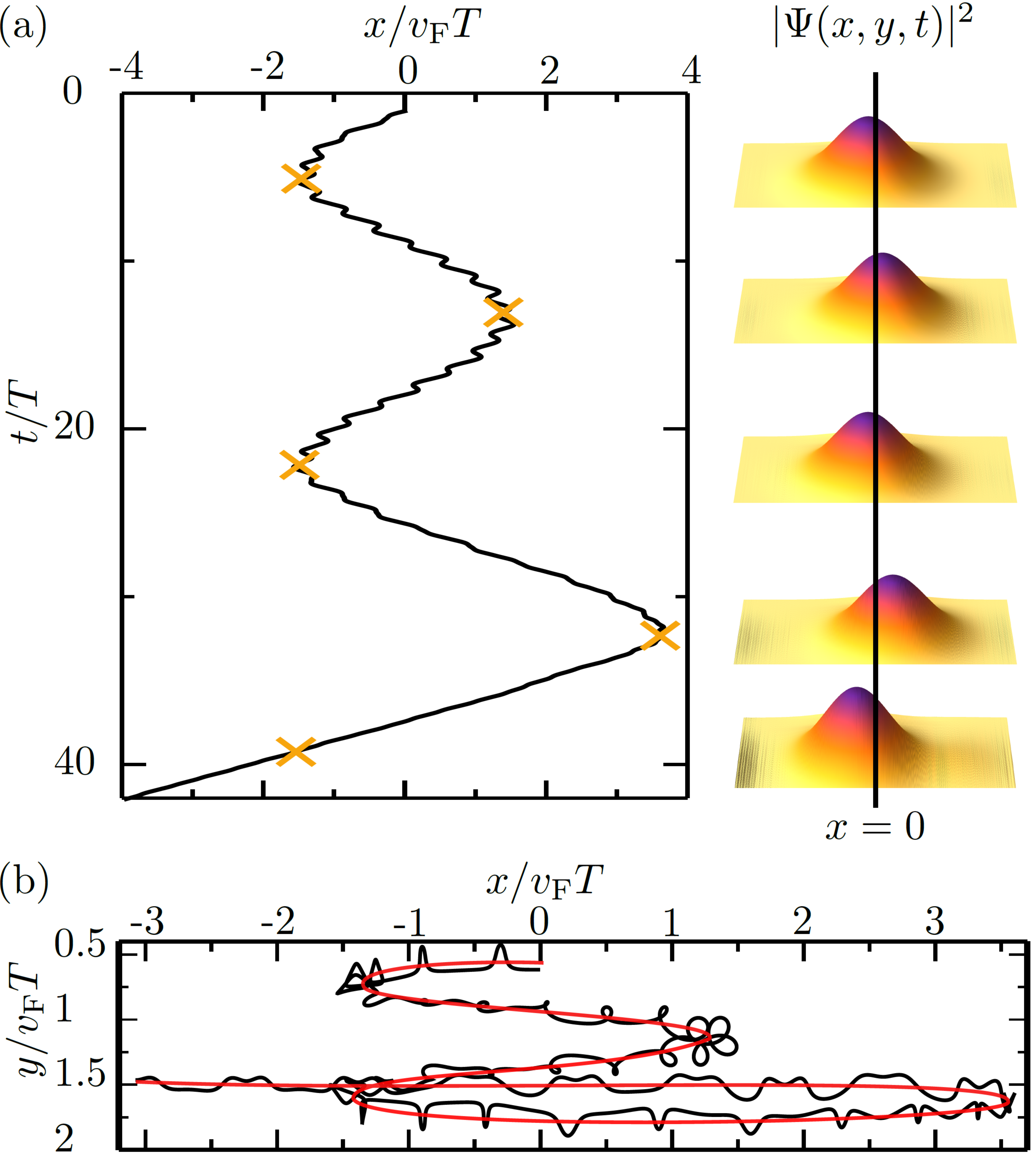}
\caption{(a) Simulation of the position expectation value of the WP in the direction of the electric field, Eq.~\eqref{eq:Efield}, (as shown in Fig.~\ref{circpollight}(b)) and its shape in position space. The orange crosses mark when the snapshots of the WP are taken. The tiny oscillations on top of the comparably slow Floquet oscillations are due to the momentum change caused by the circulating electric field of the pulse, Eq.~\eqref{eq:lightfield}. (b) Trajectory of the WP's center of mass $\langle \hat{ \bb x} \rangle (t)$ displayed in (a) in $x$-$y$-plane. The red curve has been included as a guide for the eye.}
\label{x-y_WP_circ}
\end{figure}


\subsection{Including trigonal warping and introducing experimental parameters}
\label{trigwarp}

\begin{figure}
\centering
\includegraphics[width=\columnwidth]{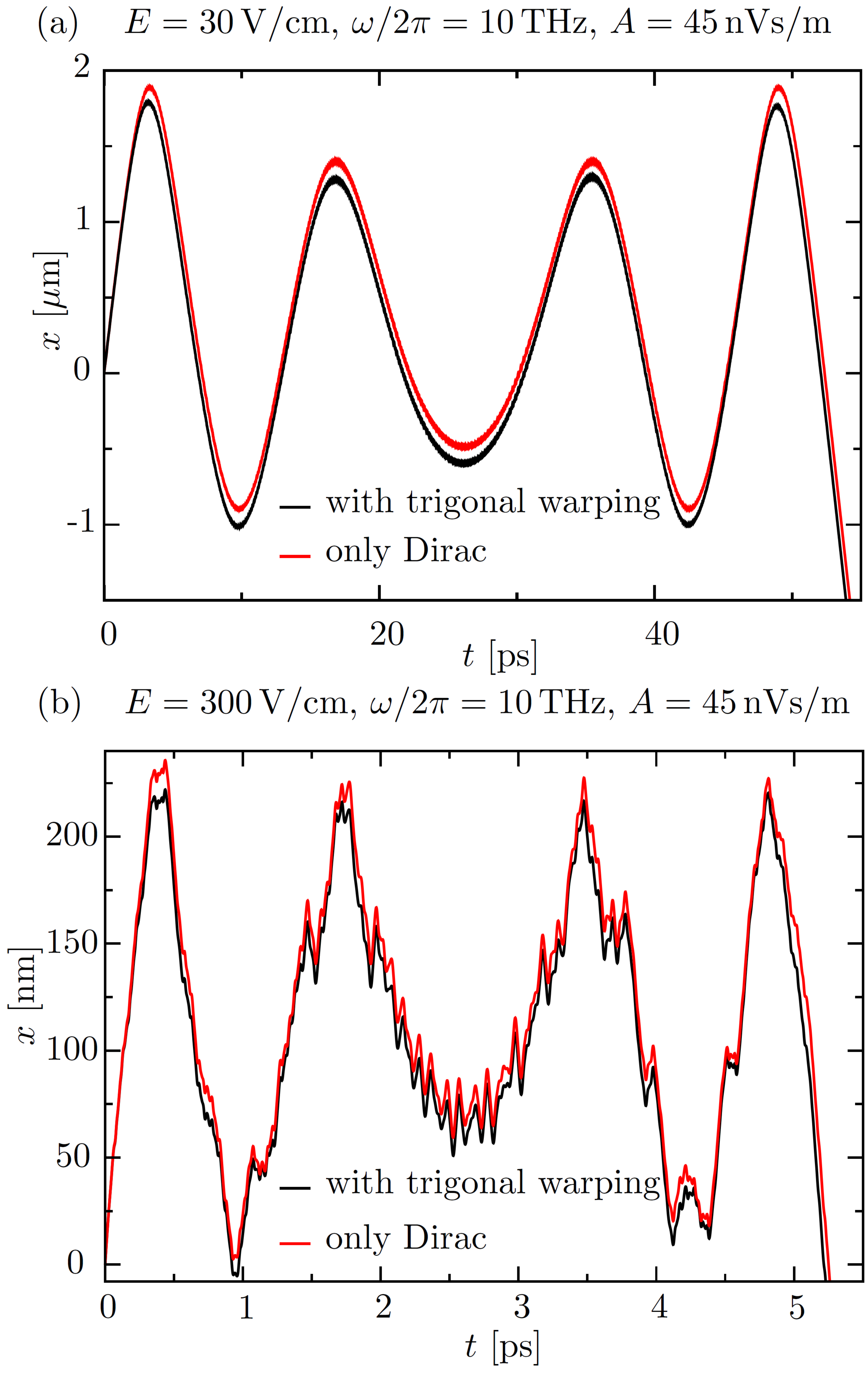}
\caption{Floquet oscillations in graphene with a driving frequency $\omega/2\pi = 10 \,\rm Thz$ and laser amplitude $A = 45 \,\rm nVs/m$. (a) When a static electric field $E = 30 \,\rm V/cm$ is applied, clear Floquet oscillations of period $T_F \simeq 13.5 \,\rm ps$ are visible. (b) Increasing the electric field $E$ by a factor of 10 reduces the Floquet period to $T_F \simeq 1.3 \,\rm ps$. Because of the smaller time-scale, the oscillations due to the momentum change induced by the circularly polarized light ($T = 0.1 \,\rm ps$) are now resolved. Their features are additionally altered by Zitterbewegung. For both cases (a) and (b) including trigonal warping in the calculations does not lead to qualitative changes of the Floquet oscillations.}
\label{warpingexperiment}
\end{figure}

Finally, we want to give a few more details on the experimental realizability of Floquet oscillations in graphene. First of all we numerically verify that when choosing the correct energy scales trigonal warping effects do not play a role. Therefore we extend our Hamiltonian $H_0(\bb k)$, Eq.~\eqref{DiracHamiltonian}, to
\begin{equation}
H(\bb k) = \hbar v_F \bb k \cdot \boldsymbol{\sigma} - \mu \left[ \left(k_y^2 - k_x^2 \right) \sigma_x + 2 k_x k_y \sigma_y \right], 
\end{equation}\\ 
with $v_F = 10^6 \,\rm m/s$ and $\mu = 3 a^2 t/8$, where $a = 1.42 \,\rm \AA$ is the nearest-neighbor distance and $t = 2.7 \,\rm eV$ the nearest-neighbor hopping strength of graphene \cite{graphenebook}. We set the light amplitude $A = 45 \,\rm nVs/m$ and the frequency $\omega/2\pi = 10 \,\rm THz$. We simulate the wave packet motion for $\bar{k}_i = 0.013 \,\rm 1/\AA$. The described parameters are equivalent to the unitless values chosen for Fig.~\ref{circpollight}. In Fig.~\ref{warpingexperiment} we compare the Floquet oscillations obtained with and without the trigonal warping term and find no qualitative differences. However, for experimental realization in graphene the transport relaxation time $\tau$ has to be higher than the period $T_F$ of the Floquet oscillations, Eq.~\eqref{eq:T-Floquet}, typically $\tau = 1 - 20 \,\rm ps$ \cite{wang2013,berdyugin2019}. As $T_F \simeq 13.5 \,\rm ps$ for $E = 30 \,\rm V/cm$ (Fig.~\ref{warpingexperiment} (a)), good quality samples would already allow for the observation of one period. If we increase the electric field by a factor of 10 (Fig.~\ref{warpingexperiment} (b)), $T_F \simeq 1.3 \,\rm ps$. Then, all 4 Floquet oscillation cycles could be detected. Note that for the higher electric field the Floquet oscillations become slightly altered. The additional oscillations are caused by Zitterbewegung and the momentum change induced by the circularly polarized light field and are resolved here since the Floquet oscillations occur on the time-scale of pico-seconds.

\section{Conclusions}

The above analysis and simulations constitute a proof of principle for generating Floquet oscillations
in systems with an effective Dirac dispersion.
Concerning possible experimental realizations, graphene \cite{GrapheneelectronicNeto},
topological insulators \cite{topInsuReviewKane,topInsuReviewZhang} and cold atoms in artificial honeycomb lattices \cite{artificialgraphene} appear suitable \cite{kolovsky2013, kolovsky2018}. The latter have the advantages that one can additionally tune $v_F$ entering the Floquet frequency and that relaxation through disorder or interaction effects can be avoided. Other effective Dirac systems, \eg for plasmons 
\cite{diracplasmons} or polaritons \cite{diracpolaritons}, could also be 
considered. In the following, we will quantitatively focus on realizations for charge carriers in real monolayer graphene. 

The radiation frequency $\omega$ must be chosen such that several Floquet BZs lie in the energy range governed by the linear dispersion. In graphene, a frequency $\nu = 2\pi\omega$ of $1-10\,$THz is small enough to accommodate roughly 50 to 5 Floquet replicas over the energy range of $400\,$ meV, for which the linear Dirac cone is a very good approximation. Additionally, $\omega > \omega_F \gg 2\pi/\tau$ is required, where $\tau$ denotes a typical relaxation time of charge carriers. To realize the oscillations shown in Fig.~\ref{warpingexperiment}(b) with $\tilde{A}=1.1$ at a radiation frequency of $\omega/2\pi = 10  \,\rm THz$, the moderate intensity (avoiding sample heating) of $I = \frac{c \epsilon_0}{2} A^2 \omega^2 \simeq 1 \frac{\rm MW}{\rm cm^2}$ is needed. Here $c$ is 
the speed of light and $\epsilon_0$ the vacuum permittivity. Then an electric field $E\simeq 0.3 \frac{\rm kV}{\rm cm}$ is sufficient to generate Floquet oscillations of frequency $\omega_F/2\pi \gtrsim 1 \, \rm THz$. Hence $\omega_F \gtrsim 2\pi/\tau $ for typical inverse transport relaxation times $1/\tau = 0.05 - 1 \, \rm THz$ of clean hexagonal boron nitride-encapsulated graphene \cite{wang2013,berdyugin2019}. Thus, Floquet oscillations could in principle be observed, opening an alternative way to generate THz radiation~\cite{ganichevbook,THzReview}.


To conclude we showed that free particles in a static electric drift field and obeying a linear Dirac-type 
dispersion can perform spatially periodic motion, Floquet oscillations, when subject to time-periodic driving. The Floquet
time lattice takes on the role of the spatial lattice required for conventional Bloch oscillations.
Such Floquet oscillations feature zitterbewegung and characteristic amplitude modulations 
that could provide a tool to experimentally map the Floquet quasi-bands. 
A closer consideration of Landau-Zener transitions between different 
Floquet bands and the question of how the topology of Floquet bands \cite{FloquetTIGalitski} is reflected in 
corresponding Floquet oscillations opens interesting perspectives for future research.

\begin{acknowledgments}
We thank Simon Maier for support with calculating Floquet band structures at an early stage of 
this work and Sergey Ganichev for helpful discussions and careful reading of the manuscript. We further thank an anonymous referee for suggesting to interprete Floquet oscillations in terms of multiple photon exchange.
We acknowledge funding by the Deutsche Forschungsgemeinschaft (DFG, German Research Foundation) -- Project-ID 314695032 -- CRC 1277 (subproject A07).
\end{acknowledgments}


\appendix

\section{Basic relations in Floquet theory}
\label{suppFloquet}

In the following we describe how to obtain the Floquet band structure for a finite driving strength $V_T(t)$ without the electric field, which is later included by a shift of $k$ when computing Floquet oscillations. 
Generally, the Floquet operator $\mathcal{H}_F$ is given by 
\begin{equation}
\mathcal{H}_F = H_0(\bb k)+V_T(t)-i\hbar \partial_t,
\label{suppeq:floquetH} 
\end{equation}
where $H_0(\bb k)$ is the Hamiltonian of the time-independent system. Since the eigenstates of $\mathcal{H}_F$, $\Phi_\alpha(\bb k,t)$, are periodic in time, they can be expanded in a Fourier series:
\begin{equation}
\Phi_\alpha(\bb k, t) = \sum_{n=-\infty}^\infty \bb u_\alpha(\bb k, n) \rme^{in\omega t}. 
\end{equation}
The dimension of the coefficients $\bb u_\alpha(\bb k, n)$ is equal to the number of branches $\alpha$ of the time-independent Hamiltonian $H_0(\bb k)$, \ie two in our case. In order to find these coefficients and the corresponding eigenenergies $\epsilon_\alpha$, the Floquet equation \eqref{suppeq:floquetH} is multiplied by $\rme^{-im\omega t}$, $m \in \mathbb{Z}$, and averaged over one period $T$ to end up with \cite{floquetshirley, floquetsambe}
\begin{equation}
\begin{split}
\sum_{n=-\infty}^{\infty}& \underbrace{\left( \mathcal{H}_{0F, mn}(\bb k) + V_{F, mn} \right)}_{\mathcal{H}_{F, mn}(\bb k)} \bb u_\alpha (\bb k, n)\\\ 
&= \epsilon_\alpha(\bb k) \bb u_\alpha (\bb k, m).
\label{suppeq:floqueteigenvalues}
\end{split}
\end{equation}
Here the contributions from $V_T(t)$ are denoted by $V_{F, mn}$ and the contributions from $H_0(\bb k)$ and $-i\hbar\partial_t$ by $\mathcal{H}_{0F, mn}(\bb k)$.
Note that the resulting eigenvalue problem is time-independent and all the dynamics of a WP in the system are incorporated in the Floquet basis states \cite{Holthaus_super_Bloch}. 
To account for this when projecting a WP from the time-dependent basis to the Floquet basis, the relation $\langle k_o(t) \rangle_t = k_{Floquet}$ is used, where $\langle k_o(t) \rangle_t = \frac{1}{T}\int_0^T k_o(t)\ \rmd t$. 
The time dependence of $k_o(t)$ is introduced by the time-periodic driving $V_T(t)$. 
Hereafter, we refer to $k_{Floquet}$ when talking of wave numbers and  suppress its subscript to simplify the notation. 
The Floquet Hamiltonian without coupling
\begin{equation}
\mathcal{H}_{0F, mn}(\bb k) = (\hbar v_F \bb k \cdot \bs\sigma + m\hbar \omega) \delta_{mn}
\end{equation}
is diagonal and describes the Dirac bandstructure shown in Fig.~\ref{floquetbandslinear}. The driving Hamiltonian
\begin{equation}
V_{F, mn} = \frac{1}{T} \int_0^T \rmd t \  V_T(t) \rme^{i(n-m)\omega t} 
\label{equationV_f}
\end{equation}
on the other hand couples different Floquet modes $\bb u_\alpha(\bb k, n)$ and thus can lead to the opening of band gaps in the originally linear spectrum (see Fig.~\ref{floquetbandslinear}). For the numerical evaluation, the resulting infinite matrix $\mathcal{H}_F$ has to be truncated at a finite value $\pm n$ that corresponds to the number of Floquet replicas taken into account. When performing numerical calculations, one has to make sure that the results are converged for the chosen value of $n$.

\section{Floquet quasi-band structure of a Dirac system with periodically opened mass gap}
\label{suppBands}
Here we give a more detailed analysis of the Floquet band structure of a Dirac system with a periodically opened mass gap. The potential describing this time-dependent pulsing is given in Eq.~\eqref{potentialmass}. 
For the mass gap, we find with Eq.~\eqref{equationV_f} 
\begin{equation}
V_{F, mn} = \frac{M}{2\pi i (n-m)} \left(1-\rme^{-2\pi i(n-m)\frac{\Delta t}{T}} \right) \sigma_z.
\end{equation}
Qualitatively, the two time scales $\Delta t$ and $T$ involved are reflected in $k$-space. 
While the longer scale $T$ makes for the high-frequency oscillations of the Floquet bands due to the replicas, the smaller time scale $\Delta t$ is responsible for the slow modulation, \ie the different gap sizes as function of $k$ shown in Fig.~\ref{floquetbandsprob}.

On a more quantitative level, the dependence of the Floquet bands on the pulse amplitude $M$ and pulse duration $\Delta t$ can be best understood when studying the influence of one pulse on a WP in the static Dirac cone. This has been done extensively in Ref.~\cite{qtmreck} and will only be summarized here. 
Let us consider a WP initially occupying states in the upper cone. The opening and closing of the gap causes a redistribution of the WP to both cones, leading to a new superposition state. The amplitude of the part now occupying the other cone - in our example the lower one - can easily be calculated analytically \cite{qtmreck}:
\begin{equation}
A(k) = -\frac{i}{\sqrt{1+\eta^2}} \sin \left(\mu \sqrt{1+ \eta^2} \right), 
\end{equation} 
where $\eta = E_{k, \pm}/M$ and $\mu = M \Delta t/\hbar$. This transition amplitude only depends on the initial energy $E_{k, \pm}$ of the state and the pulse parameters. 
\begin{figure}
\centering
\includegraphics[width=\columnwidth]{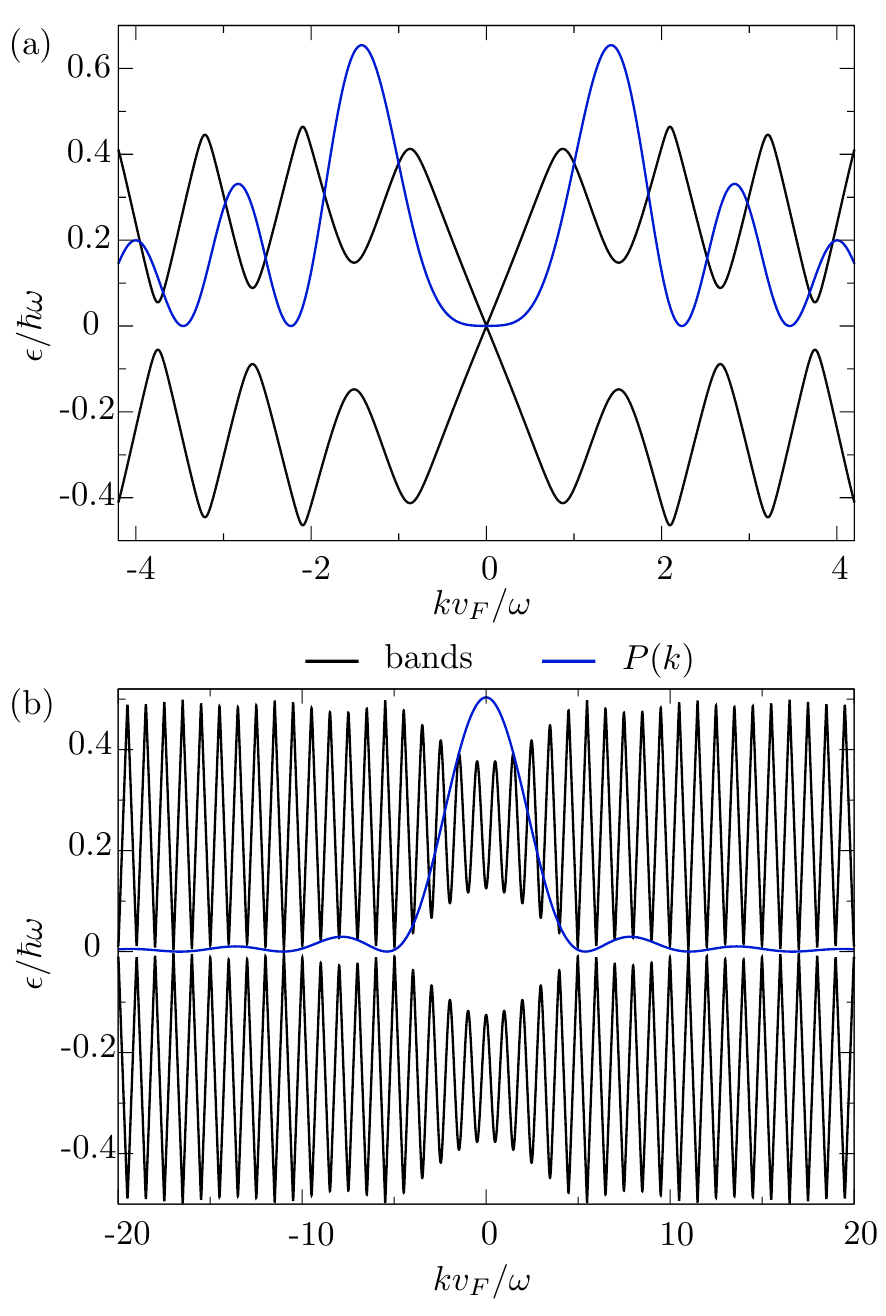}
\caption{Floquet band structure (black curves) and corresponding transition probability $P(k)$ (blue) for (a) $\mu = 6.3$, $\Delta t/T = 0.5$ and (b) $\mu = 0.4$, $\Delta t/T=0.09$. The bands shown in Fig.~2(a) of the main paper are a zoom into those depicted in panel (b). For both parameter sets one can easily see that the gap in the Floquet bands closes whenever $P(k)$ goes to zero.}
\label{floquetbandsprob}
\end{figure}
A numerical comparison of the corresponding transition probability $P(k)=|A(k)|^2$ and the Floquet band structure reveals that the gaps that open at the intersections of the repetitions of the original cone are directly related to the transition probability at that $k$-value: The larger $P(k)$, the larger the band gap.
The reason for this dependence can be motivated in the following way.
$P(k)$ describes for a single pulse the proportion of the WP that is transferred to the other band, \ie how much a single pulse couples upper and lower band of the Dirac cone.
On the other hand, the band gap of the Floquet bands is due to this coupling of (initially) linear band replicas.
Therefore, it is not surprising that $P(k)$ and the band gap width are directly related.

We show this for two exemplary band structures in Fig.~\ref{floquetbandsprob}. In panel (a) we set $\mu = 6.3$ and $\Delta t/T = 0.5$. Since $P(k)=0$ around $kv_F/\omega=0$ the original Dirac cone is preserved. For larger $k$ band gaps open. 
The resulting complex band structure is a perfect example of how nicely the Floquet band structure can be tuned based on the transition probability $P(k)$. The band structure shown in panel (b) is the same as the one shown in Fig.~2(a) of the main paper bur for a larger range of $k$-values. 
There, a wide area around $kv_F/\omega = 0$ is gapped. 
For an appropriate choice of parameters (as in panel (b)) a large $k$-window, in which Landau-Zener transitions are suppressed, can be chosen to support Floquet oscillations.
As a rule of thumb, the smaller $\Delta t/T$, the more band gaps open and thus allow for more periods of Floquet oscillations before Landau-Zener transitions diminish them.

\section{Zitterbewegung in a Dirac system with periodically opened mass gap}
\label{suppZB}

\begin{figure}
\centering
\includegraphics[width=\columnwidth]{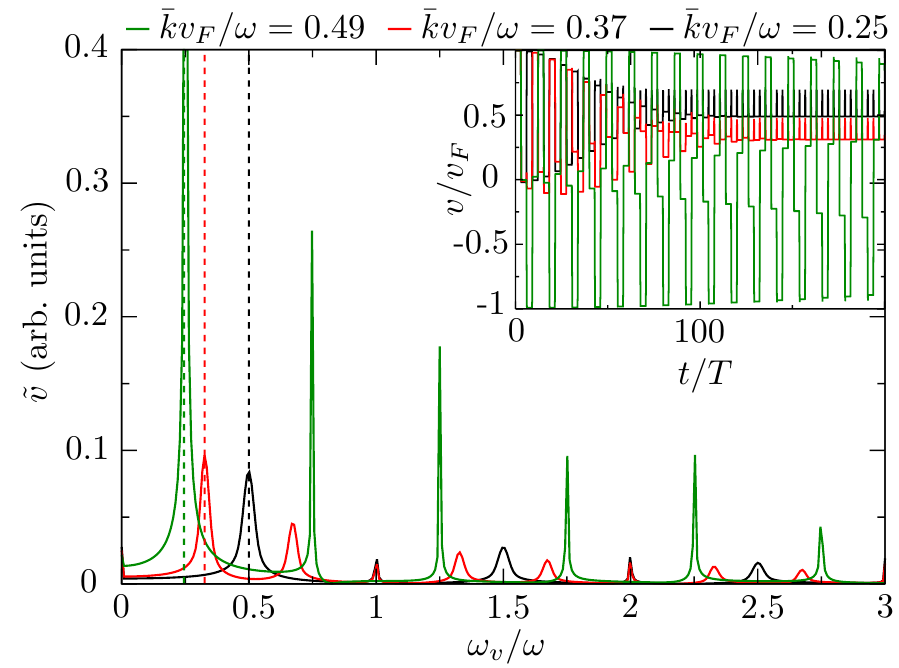}
\caption{Frequency analysis by a Fourier transform of the velocity $\tilde{v}(\omega_v)$ of WPs with different $\bar{k}$. 
Since no electric field is applied, $k$ is constant. 
The dashed lines mark $\omega_{ZB}$ as expected from Eq.~\eqref{omegaZB} which match perfectly the numerically obtained peaks. 
The inset shows the corresponding time-dependency of the velocities. For the Fourier analysis, the respective average velocities have been subtracted to avoid a zero-frequency peak.}
\label{frequency_ZB}
\end{figure}

Zitterbewegung (ZB) was originally predicted by Schr\"odinger for the Dirac equation \cite{schroedinger1930zitterbewegung} but an analogue is also visible in multiband systems \cite{schliemann2005,gerritsma2010,zawadzki2011,qu2013,leblanc2013,ZBPhillipp}.
The reason is the interference of particle and antiparticle contributions in a WP, or respectively the contributions of different bands, in cases when the velocity operator does not commute with the Hamiltonian.
The corresponding term is the offdiagonal term of the expectation value of the velocity operator and the frequency is given by the difference of the energies.
In our case, we also have an effective two band model due to the Floquet bands with an offdiagonal term of the velocity as seen in Eq.~\eqref{floquetvelocity}.
The frequency of the corresponding ``Floquet zitterbewegung'' is given by the energy gap between the two Floquet quasi-energies,
\begin{equation}
\omega_{ZB} = \left( \epsilon_{\beta}(k)-\epsilon_{\alpha}(k) \right)/\hbar.
\label{omegaZB}
\end{equation}

In Fig.~\ref{frequency_ZB} we show the Fourier transform $\tilde{v}(\omega_v)$ of the velocity of a WP starting at different $k$-values in the Floquet band structure to analyze the frequency spectrum.
We denote the variable of the Fourier transform of the velocity by $\omega_v$. 
For these calculations no static electric field was applied, such that $k$ and therefore $\omega_{ZB}$ stay constant in time. 
Before performing the Fourier transform to investigate the frequency of the oscillations, the mean velocity value was subtracted of the corresponding data to avoid a peak at $\omega_{v} = 0$.

The dashed lines mark $\omega_{ZB}$ as calculated by Eq.~\eqref{omegaZB}. Their good agreement with the spectrum confirms that the off-diagonal velocity in the Floquet picture describes ZB caused by the interference of states occupying different Floquet bands. Since the velocity has a rectangular shape, the peaks are repeated at higher harmonics.  
This rectangular shape can be explained in our example by the fact that the velocity can only change during the mass gap, which means that the harmonic oscillation of the ZB is effectively sampled with the driving frequency $\omega$.
%

\section{The c++ library "Time-dependent Quantum Transport" (TQT)}
\label{suppTQT}

To propagate a quantum state $|\psi\rangle$, one has to solve the time-dependent Schr\"odinger equation,
\begin{equation}
 \rmi \hbar \frac{\partial}{\partial t} |\psi\rangle = \hat H |\psi \rangle,
\label{eq:genSchroe}
\end{equation}
with the Hamilton operator $\hat H$, which depends in general on time.
Formally, it can be solved using the  time-evolution operator 
\begin{equation}
 U(t,t_0) = \mathcal{T}\exp\left(- \frac{\rmi}{\hbar} \int_{t_0}^{t}\hat H(t^\prime) \,\rmd t^\prime\right),
\label{eq:Ut-ordered}
\end{equation}
which is unitary and fulfills 
\begin{equation}
  U(t, t_0) =  U(t, t^\prime)  U(t^\prime, t_0),
\end{equation}
where $t_0$ is the initial time, $t$ is some arbitrary later time and $t^\prime$ is a time in between.
The time-evolution of a state then yields
\begin{equation}
 |\psi(t) \rangle = U(t, t_0) |\psi(t_0)\rangle.
\end{equation}
Moreover, for time-independent Hamiltonians, the time-evolution operator simplifies to 
\begin{equation}
 U(t_0, t) = \exp\left(-\rmi \frac{\hat H}{\hbar}  \cdot (t-t_0)\right).
\end{equation}
On the other hand, any function can be approximated by step-wise constant functions -- the smaller the steps, the better the approximation.
Thus, the time-ordered exponential of Eq.~\eqref{eq:Ut-ordered} can be estimated by
\begin{equation}
 U(t_0, t_0+ N\delta t) \approx \prod\limits_{j=0}^{N-1} \exp\left(-\rmi \frac{\hat H(t_0+j\delta t)}{\hbar}  \cdot \delta t\right),
\end{equation}
where the Hamiltonian is made step-wise constant for the time duration $\delta t$. 
The advantage is that instead of the time-ordered product of Eq.~\eqref{eq:Ut-ordered}, a rather easy multiplication can be performed. 
Of course, one has to make sure that the time step $\delta t$ is small enough, such that the numerical result is converged.

Using the time-evolution operator, we shifted the problem of solving the differential equation in Eq.~\eqref{eq:genSchroe}, to having the Hamiltonian operator in an exponential, which is defined by its (infinite) series expansion.
The publicly available c++ library ``Time-dependent Quantum Transport'' (TQT) by Viktor Krueckl \cite{phdkrueckl} takes care of this expansion as efficient as possible for 1d or 2d systems.
The expanded time-evolution operator acts on a numerically defined initial state in real space.
Since an sufficiently smooth function can be approximated by its values at discrete points, the space is discretized by a grid, in 1d with $N_x$ and in 2d with $N_x\times N_y$ points. 
Thus, the wave function becomes a complex valued $N_x$-component vector or $(N_x\times N_y)$-matrix, respectively. For this work we usually take $N_x = 8192$ for 1d and $N_x \times N_y = 8192\times256$ for 2d systems.

The Hamiltonian can be either given as tight-binding Hamiltonian or as mixed position and momentum space representation, \ie a function of both, the position and momentum operator, the latter being the Hamiltonian used in most cases  for analytical calculations. 
In the mixed representation, instead of using the spatial derivative, the momentum operator acts in momentum space, \ie the wave function is transformed by a fast Fourier transform, then the momentum operator acts as factor, and finally the inverse fast Fourier transformation is applied to get back to position space.
The reason for using the Fourier transformation instead of the derivative is the numerical instability of the latter. 
Since the momentum operator acts several times (in higher orders $k_i^n$) in each small time step, the errors add up quickly.
In this paper, only the mixed representation of position and momentum operator is used.

Due to the explicitly time-dependent Hamiltonian in our problem, a Lanczos method is used to expand the time-evolution operator \cite{nauts1983,park1986} instead of a Chebyshev expansion\cite{talezer1984,kosloff1994}.
The difference here is that instead of expanding in a fixed set of polynomials, the time-evolution operator is expanded in terms of the wave function $\psi$ itself and powers of the Hamiltonian acting on the wave function $\hat H^n\psi$. 
The thereby spanned subspace is a $N$-dimensional Krylov subspace $\mathcal{K} = \operatorname{span}\lbrace \psi, \hat H\psi,\dots{}, \hat H^{N-1}\psi\rbrace$, which is orthonormalized to get the basis vectors $u_n$ by a Gram-Schmidt procedure during the recursive creation for better numerical stability:
\begin{align}
u_0 &= \frac{\psi(t_0)}{|\psi(t_0)|}, \\
u_1 &= \frac{\hat H u_0 - \alpha_0 u_0}{\beta_0}, \\
u_{n+1} &= \frac{\hat H u_n - \alpha_n u_n - \beta_{n-1}u_{n-1}}{\beta_n},
\end{align}
with the overlaps $\alpha_n = \langle u_n\mid \hat H \mid u_n\rangle$ and $\beta_{n-1} = \langle u_{n-1} \mid \hat H \mid u_n \rangle$.
Note that  $u_n$ is a linear combination of powers of $\hat H$ acting on $\psi$, with highest order $n$.

The truncated Hamiltonian in this subspace becomes tridiagonal
\begin{equation}
 H_\mathcal{K} = \begin{pmatrix}
                 \alpha_0 & \beta_0 & 0 & \cdots{} & 0 \\
                 \beta_0 & \alpha_1 & \beta_1 & & 0 \\
                 0 & \beta_1 & \alpha_2 &  & 0 \\
                 \vdots{} &  &   & \ddots{} & \beta_{N-2} \\
                 0 & \cdots{} & 0 & \beta_{N-2} & \alpha_{N-1} \\
                \end{pmatrix},
\end{equation}
which can be diagonalized by conventional algorithms and enables the calculation of approximate eigenvalues of the operator $\hat H$ \cite{lanczos}. With the matrix of eigenvectors $\mathbf{T}$ and eigenvalues $\mathbf{E}$ of the Hamiltonian in the reduced Krylov space $H_\mathcal{K}$, the time-evolution of one small time step is given by
\begin{equation}
\psi(t+\delta t) = \sum\limits_{n=0}^{N-1} \left\lbrack \mathbf{T}^t \exp\left( -\frac{\rmi}{\hbar} \mathbf{E} \delta t \right) \mathbf{T} \;\psi_\mathcal{K}(t)\right\rbrack_n \cdot u_n.
\end{equation}
The expansion in the Krylov subspace is faster than a Taylor expansion \cite{hochbruck1997} and for the Krylov space, a dimension $N$ in the range $10$--$40$ is usually enough.
It turned out that for the calculations in this paper, the dimension of the Krylov space of $N=15$ is sufficient.

With the thus obtained time-dependent state on our discrete timeline, an arbitrary (observable) quantity like the position expectation value can be obtained as a function of the time for the propagation, which yields in our case the Floquet oscillations.

\vspace*{\fill}


%

\cleardoublepage

\end{document}